\documentclass[11pt]{article} \usepackage{mystyle-new}
\usepackage{epsfig,amsmath} \usepackage{graphics}
\usepackage{cite,./mcite}
\usepackage{subfigure} 
  {\begin{list}{}{\settowidth{\labelwidth}{#1}%
  \setlength{\leftmargin}{\labelwidth}%
  \addtolength{\leftmargin}{\labelsep}%
  \setlength{\itemsep}{0pt plus 1pt}
  \setlength{\parsep}{0pt plus 1pt}
  \setlength{\topsep}{0pt plus 1pt}
  \setlength{\partopsep}{0pt plus 1pt}
  \setlength{\parskip}{2mm plus 1mm minus 1mm}
  }}%
  {\end{list}}

\def\cascade{{\sc Cascade}}

\def\MGvATNLO{{\sc MadGraph5\_aMC@NLO}}

\def\lsim{\mathrel{\rlap{\lower4pt\hbox{\hskip1pt$\sim$}}
    \raise1pt\hbox{$<$}}}                
\def\gsim{\mathrel{\rlap{\lower4pt\hbox{\hskip1pt$\sim$}}
    \raise1pt\hbox{$>$}}}                
\def\PZ{\ensuremath{\rm{Z}}}
\def\Pg{\ensuremath{\rm{g}}}

\def\Paqu{\ensuremath{\bar{\rm{u}}}}
\def\Pqb{\ensuremath{\rm{b }}}
\def\Paqb{\ensuremath{\bar{\rm{b }}}}
\def\Pqc{\ensuremath{\rm{c}}}

\def\Pl{\ensuremath{\rm{l}}}
\def\Pal{\ensuremath{\bar{\rm{l }}}}
\def\Pp{\ensuremath{\rm{p }}}
\def\Pap{\ensuremath{\bar{\rm{p }}}}
\def\TeV{\ensuremath{{\rm{TeV }}}}
\def\GeV{\ensuremath{{\rm{GeV }}}}

\def\kt{\ensuremath{k_{\rm T}}}

\def\pt{\ensuremath{p_{\rm T}}}

\newcommand{\alphas}{\ensuremath{\alpha_\mathrm{s}}}

\newcommand{\PBM}{PB}

\newenvironment{tolerant}[1]{\par\tolerance=#1\relax}{ \par }
\newcommand{\fourfl}{4FLVN}
\newcommand{\fivefl}{5FLVN}

\newcommand{\dglap}{Gribov:1972ri,Lipatov:1974qm,Altarelli:1977zs,Dokshitzer:1977sg}

\usepackage{lineno}

\begin{document}

\begin{flushright}
DESY 21-094\\
16 June 2021
\end{flushright}
\begin{center}
 {\sffamily\Large\bfseries 
 TMD parton densities and corresponding parton showers:
 the advantage of four- and five-flavour schemes
\\ \vspace*{0.2cm}
}

{ 
 \Large 
H.~Jung, 
S.~Taheri~Monfared
}
\\  \vspace*{0.2cm} {\large Deutsches Elektronen-Synchrotron DESY, Germany}
\end{center}

\begin{abstract}
\noindent
The calculations of $\PZ + \Pqb\Paqb$ tagged jet production performed in the four- and five-flavour schemes allow for detailed comparison of the heavy flavour structure of collinear and  transverse momentum dependent (TMD) parton distributions as well as for detailed investigations of heavy quarks radiated during the initial state parton shower cascade.

We have determined the first set of collinear and TMD parton distributions in the four-flavour scheme with NLO DGLAP splitting functions within the Parton-Branching (\PBM) approach. The four- and five-flavour \PBM -TMD distributions were used to calculate $\PZ + \Pqb\Paqb$ tagged jet production at LHC energies and very good agreement with measurements obtained at $\sqrt{s} = 8, 13 $ \TeV\ by the CMS and ATLAS collaborations is observed.

The different configurations of the hard process in the  four- and five-flavour schemes allow for a detailed investigation of the performance of heavy flavor collinear and TMD parton distributions and the corresponding initial TMD parton shower, giving confidence in the evolution of the \PBM -TMD parton densities as well as in the \PBM -TMD parton shower.

 \end{abstract}

\section{Introduction} 
\label{Intro}

Practical theoretical predictions for experimental measurements at high-energy hadron colliders require detailed simulations with the help of Monte Carlo event generators including parton showers and hadronization \cite{Sjostrand:2014zea,Bahr:2008pv,Bellm:2015jjp,Bothmann:2019yzt,Baranov:2021uol}. While significant progress has been achieved in the last decade on the simulation of processes by matching and merging methods~ \cite{Frixione:2007vw,Hamilton:2012np,Hoeche:2011uq,Alwall:2014hca,Frederix:2012ps}  to combine next-to-leading order (NLO) matrix element calculations with parton showers, the parton shower is still treated separately from the parton densities. Only the Parton-Branching (PB) approach~\cite{Martinez:2018jxt,Hautmann:2017fcj,Hautmann:2017xtx} with transverse momentum dependent (TMD) parton distributions allows a direct mapping of the parton shower to the TMD parton distribution. The CASCADE3 Monte Carlo event generator~\cite{Baranov:2021uol} uses \PBM -TMDs and simulates corresponding PB-TMD parton shower for initial state partons.

The description of the Drell-Yan (DY) transverse momentum spectrum at different center-of-mass energies $\sqrt{s}$ and for different DY masses  $m_{DY}$ applying \PBM -TMDs is very successful, as shown in detail in Refs.~\cite{Martinez:2020fzs,Martinez:2018jxt,Martinez:2019mwt}. In this article we describe a detailed investigation of the corresponding \PBM -TMD parton shower by making use of \PZ\ production in association with  \Pqb -flavored jets.

\begin{tolerant}{1500}
The production of b-flavor jets can be calculated in two different approaches: the four-flavor-variable-number (\fourfl ) or five-flavor-variable-number (\fivefl ) scheme. In the four flavor approach, the \Pqb -quark is only produced in the hard process, no \Pqb -quark is present in the parton density. In contrary, in the five-flavor approach, the \Pqb -quark is treated similarly as other light quarks, the mass is taken into account as a threshold, only for scales above $m_{\Pqb}$ the \Pqb -quark is included. Correspondingly, the calculation of the hard process is different: in the four-flavor scheme, the heavy quark is treated massive, while in the five-flavor scheme, except the mass threshold, all quarks are treated similarly. 
\end{tolerant}

\begin{figure}[htb]
\centering
\subfigure[$\PZ +\Pqb\Paqb$ \fourfl\ diagram]{
\includegraphics[width=0.3\textwidth]{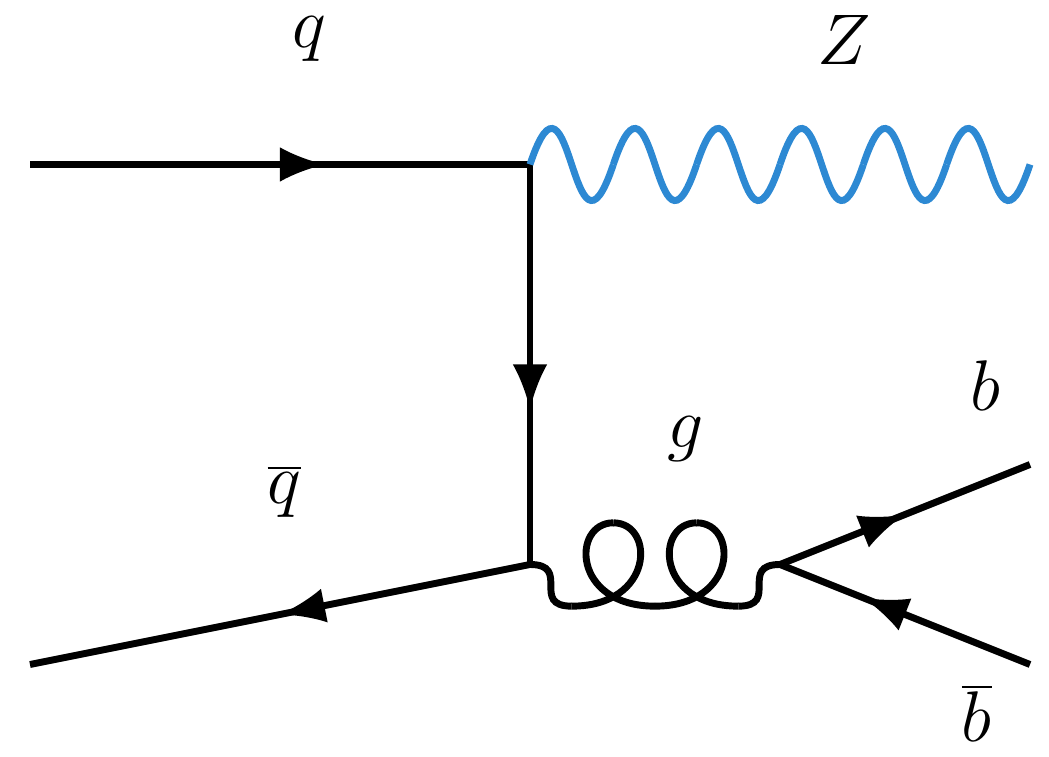}}
 \hspace{2cm}
\subfigure[$\PZ +\Pqb$ \fivefl\ diagram]{
\includegraphics[width=0.3\textwidth]{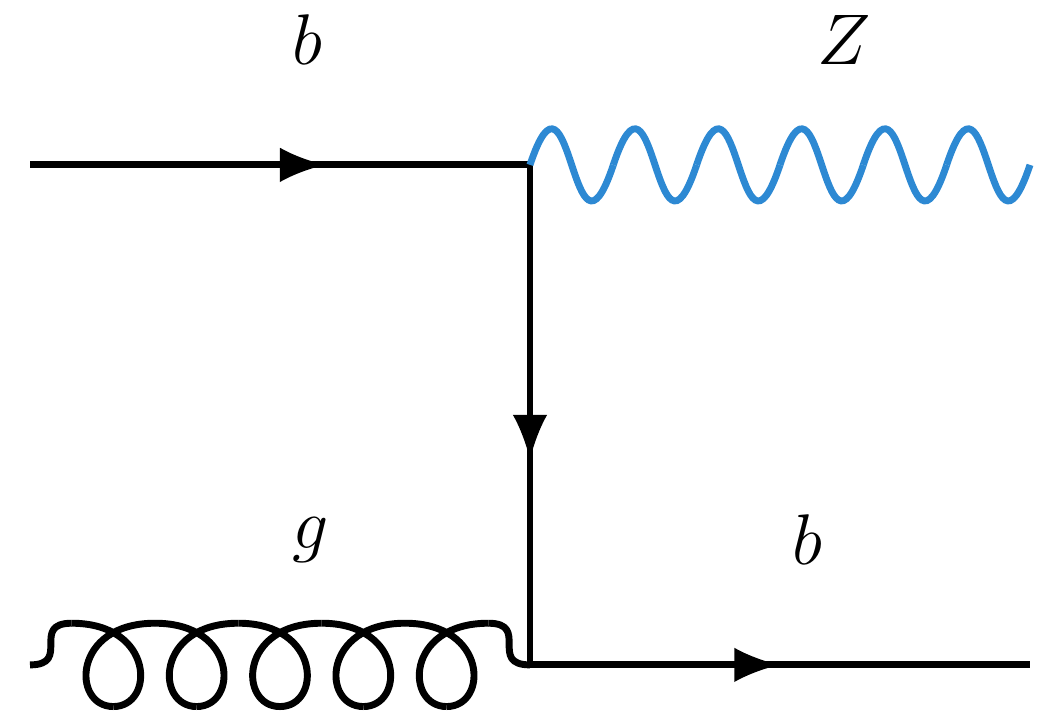}}
\caption{Example of the lowest order  diagrams for $\PZ +\Pqb$  production in \protect\fourfl\ (a) and \protect\fivefl\  (b) schemes.}\label{feynman-d}
\end{figure}

In Fig.~\ref{feynman-d} we show the lowest order process for \PZ + \Pqb -jet production at LHC energies for the \fourfl\ and \fivefl\ schemes. 

While the parton densities in the \fourfl\ and \fivefl\ schemes are different (and have to be determined separately), also the contribution from the  parton shower is different: for $\PZ +\Pqb\Paqb$ production at lowest order in the \fourfl\ scheme the matrix element plays the dominant role, in the \fivefl\ scheme the additional \Paqb -quark is contained in the parton density and is generated in the corresponding parton shower. Predictions obtained in the \fourfl - and \fivefl -schemes allow a comparison of the treatment  \Pqb\ partons in collinear and TMD parton densities as well as in the corresponding initial state TMD parton shower.

This article is organised as follows: in Section~\ref{Sec2} we describe the determination of the \fourfl\ and \fivefl\ TMD distributions applying the \PBM -method to fit inclusive HERA DIS measurements. In Section~\ref{Sec3} we describe the calculation of \PZ + \Pqb\ jet production at the LHC and compare predictions obtained in the \fourfl\ and \fivefl\ with measurements from CMS at $\sqrt{s}=8$ TeV \cite{Khachatryan:2016iob}, 
discuss the role of the \PBM -TMD parton shower and show the consistency of the \fourfl\ and \fivefl\ approaches. In Section~\ref{Sec5} we summarize our results.

\section{\PBM -TMD parton densities in  \fourfl - and \fivefl -schemes \label{Sec2}}

The \PBM\ approach was developed in Ref.\cite{Hautmann:2017fcj,Hautmann:2017xtx} to solve  the  DGLAP~\cite{\dglap} evolution equation and to calculate the TMD densities for all flavors over a large range in longitudinal momentum fraction $x$ and the evolution scale $\mu^2$. Collinear and TMD parton densities were obtained in Ref.~\cite{Martinez:2018jxt} using NLO DGLAP splitting functions, where the parameters of  the initial starting distributions were obtained from a fit to inclusive DIS measurements at HERA~\cite{Abramowicz:2015mha}.
Those parton distributions were obtained in the \fivefl -scheme using a mass of the \Pqb -quark of $m_{\Pqb} = 4.5$ \GeV\  and a value for the strong coupling at five flavors of $\alphas(m^{(5)}_{\PZ}) = 0.118$.

In the \PBM\  approach the TMD evolution equations are written as:
\begin{eqnarray}
\label{evoleqforA1}
   { {\cal A}}_a(x,\kt^2, \mu^2) 
 &=&  
 \Delta_a (  \mu^2  ) \ 
 { {\cal A}}_a(x,\kt^2,\mu^2_0) +  \sum_b 
\int
{{d { q}^{\prime 2} } 
\over { { q}^{\prime 2} } }
 \ 
 {{d \phi}
 \over
 {2\pi}
 }
{
{\Delta_a (  \mu^2  )} 
 \over 
{\Delta_a (  { q}^{\prime 2}  
 ) }
}
\ \Theta(\mu^2-{ q}^{\prime 2}) \  
\Theta({ q}^{\prime 2} - \mu^2_0)\times
 \nonumber\\ 
&&  
\int_x^{z_M} {{dz}\over z} \;
P_{ab}^{(R)} (\alphas 
,z) 
\;{ {\cal A}}_b\left({x \over z}, \kt^{\prime 2} , 
{ q}^{\prime 2}\right)  
  \;\;  ,     
\end{eqnarray}
where $ { {\cal A}}_a(x,{ \kt^2}, \mu^2)$ is the probability distribution for flavour $a$ carrying the longitudinal momentum fraction $x$ of the hadron's momentum and transverse momentum ${\kt^2}$ at the evolution scale $\mu^2$; $z$ and $q^{\prime 2}$ are the branching variables, with $z$ being the longitudinal momentum transfer at the branching, and $q^{\prime 2}$ the momentum scale at which the branching occurs; $\kt^{\prime}=|{\bf k}+(1-z) {\bf q}^\prime|$ is the rescaled transverse momentum of the emitted parton; $\phi$ is the azimuthal angle between ${\bf q}^{\prime}$ and ${\bf k}$; $\mu_0$ is the initial evolution scale; $\Delta_a$ is the Sudakov form factor; $P_{ab}^{(R)}$ are the real emission DGLAP splitting kernels.

Since the momenta ${\bf q} = {\bf q}^\prime (1-z)$ of the emitted partons are treated explicitly, color coherence effects can be imposed by using angular ordering conditions, motivating to use the transverse momentum  $q_{\rm T}=\sqrt{{\bf q}^2} = \sqrt{{\bf q}^{\prime\,2 }(1-z)^2} $ as a scale choice for $\alphas$; this condition was already applied in the PB-NLO-2018 Set 2 parton distributions described in Ref.~\cite{Martinez:2018jxt}, which is the distribution in the \fivefl -scheme.

\subsection{\fourfl\ scheme parton distributions}

In the \fourfl\ scheme, the \Pqb -quark does not appear as an active flavor in the parton evolution (practically we send $m_{\Pqb} \to \infty$)). However, care has to be taken for the value of $\alphas(m_{\PZ})$, we use $\alphas(m^{(4)}_{\PZ}) =0.1128$ as the four-flavor $\alphas$, by a prescription motivated in  Ref.~\cite{Dulat:2015mca}.

The functional form  of the starting distribution is the same as for the \fivefl\ parton distributions in Ref.~\cite{Martinez:2018jxt}, however the
 parameters of the starting distribution are re-fitted to inclusive DIS measurements from HERA~\cite{Abramowicz:2015mha} (the same data set as used for \fivefl\ parton distributions),
 yielding a $\chi^2/dof=1.254$, very close to the one obtained for the \fivefl -fit (using the \verb+xFitter+ package \cite{Alekhin:2014irh}).
 The uncertainties of the parton distributions coming from the experimental uncertainties of the data used are obtained with the Hessian method, as described in Ref.~\cite{Martinez:2018jxt}. The model uncertainties are evaluated separately by varying the values  of  $\mu_0$, $q_{cut}$ and $m_{\Pqc}$ around their default value as described in Ref.~\cite{Martinez:2018jxt}.

\begin{figure}[htb]
\subfigure [$\bar{u}$]
{\includegraphics[width=5cm]{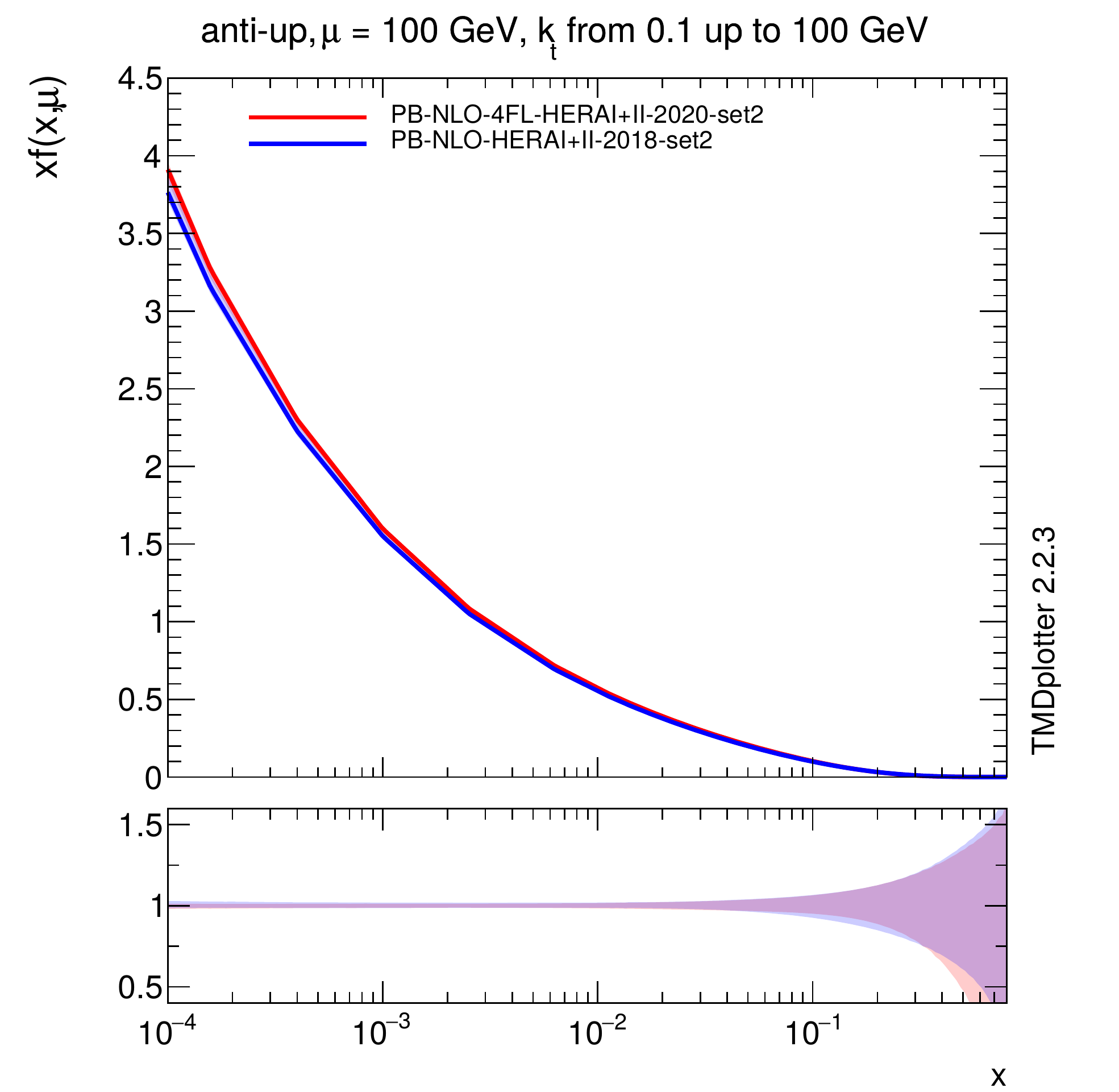}}
\subfigure[charm and bottom]
{\includegraphics[width=5cm]{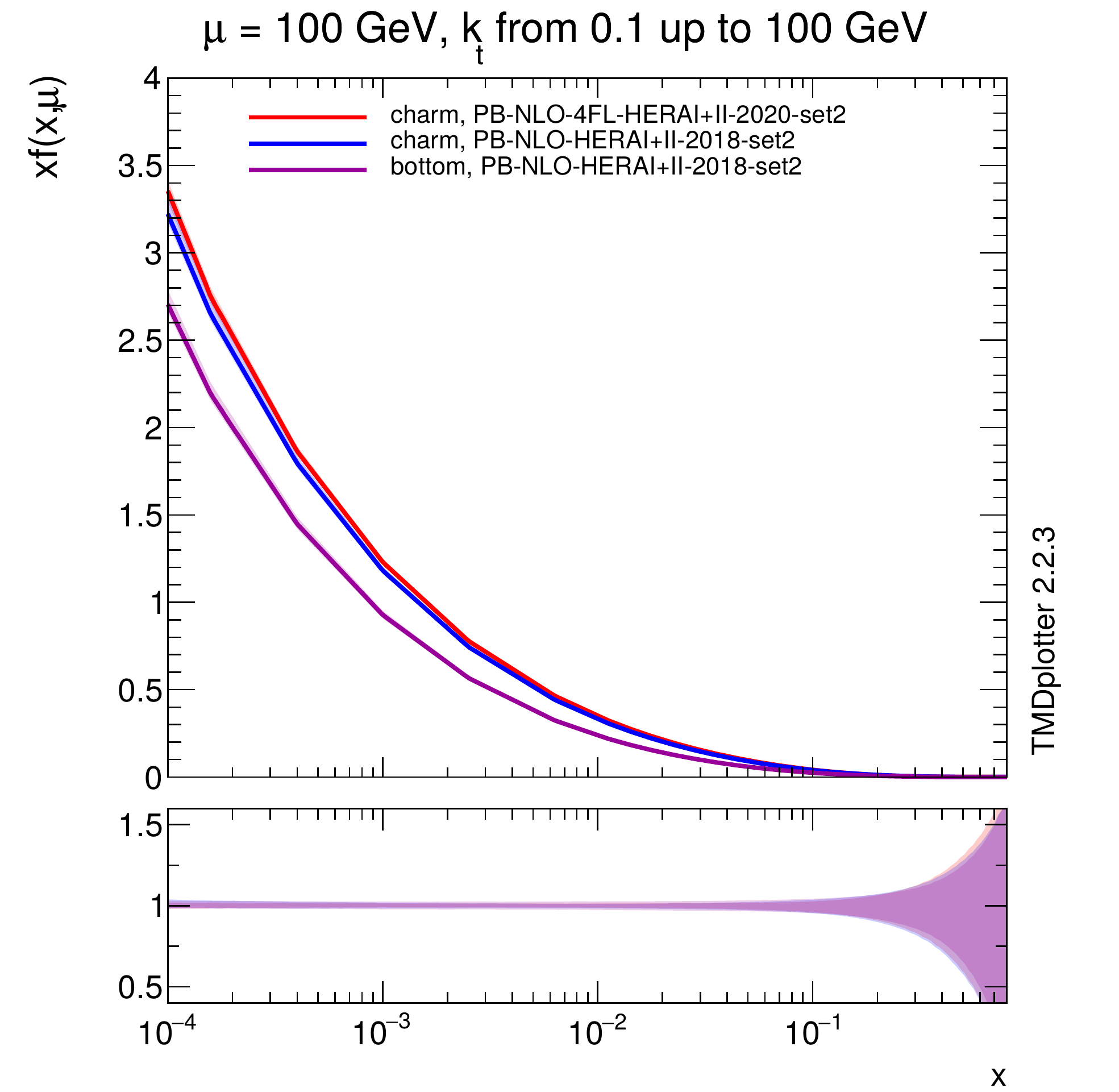}}
\subfigure[gluon]
{\includegraphics[width=5cm]{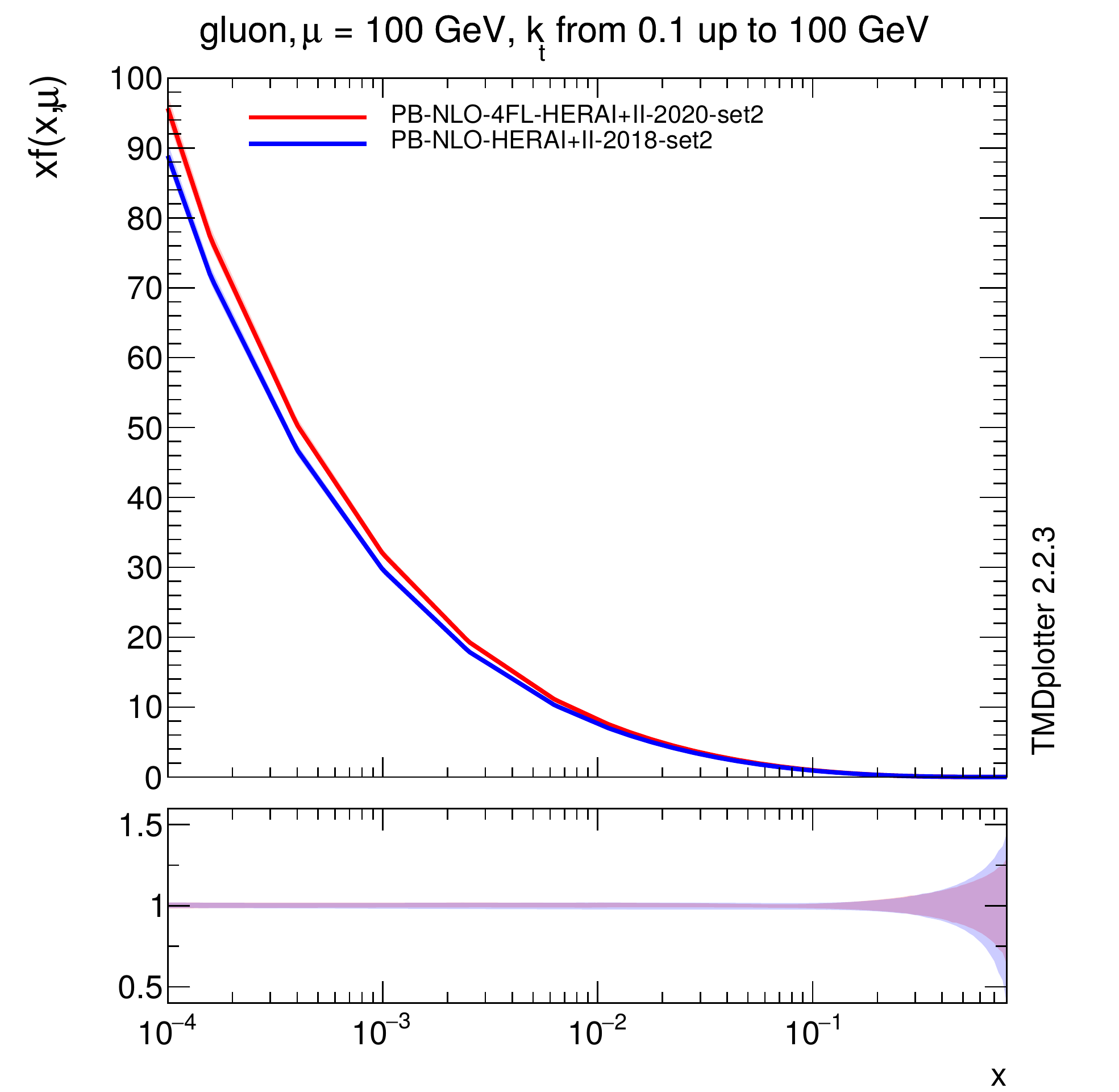}}
\caption{The $\bar{u}$ (a), charm and bottom (b) and gluon (c) \fourfl\  and \fivefl\ collinear parton densities for $\mu=100$ GeV. }\label{pdfs}
\end{figure}

In Fig.~\ref{pdfs} the \fourfl\ and \fivefl\ collinear parton densities are shown as a function of $x$ at the evolution scale of $\mu=100$ \GeV. 

Since there are less active flavors in the \fourfl -scheme, the parton density distributions are slightly higher, which is especially visible in the gluon distribution. 

\begin{figure}[htb]
\begin{center} 
\subfigure[$\bar{u}$]{\includegraphics[width=0.32\textwidth]{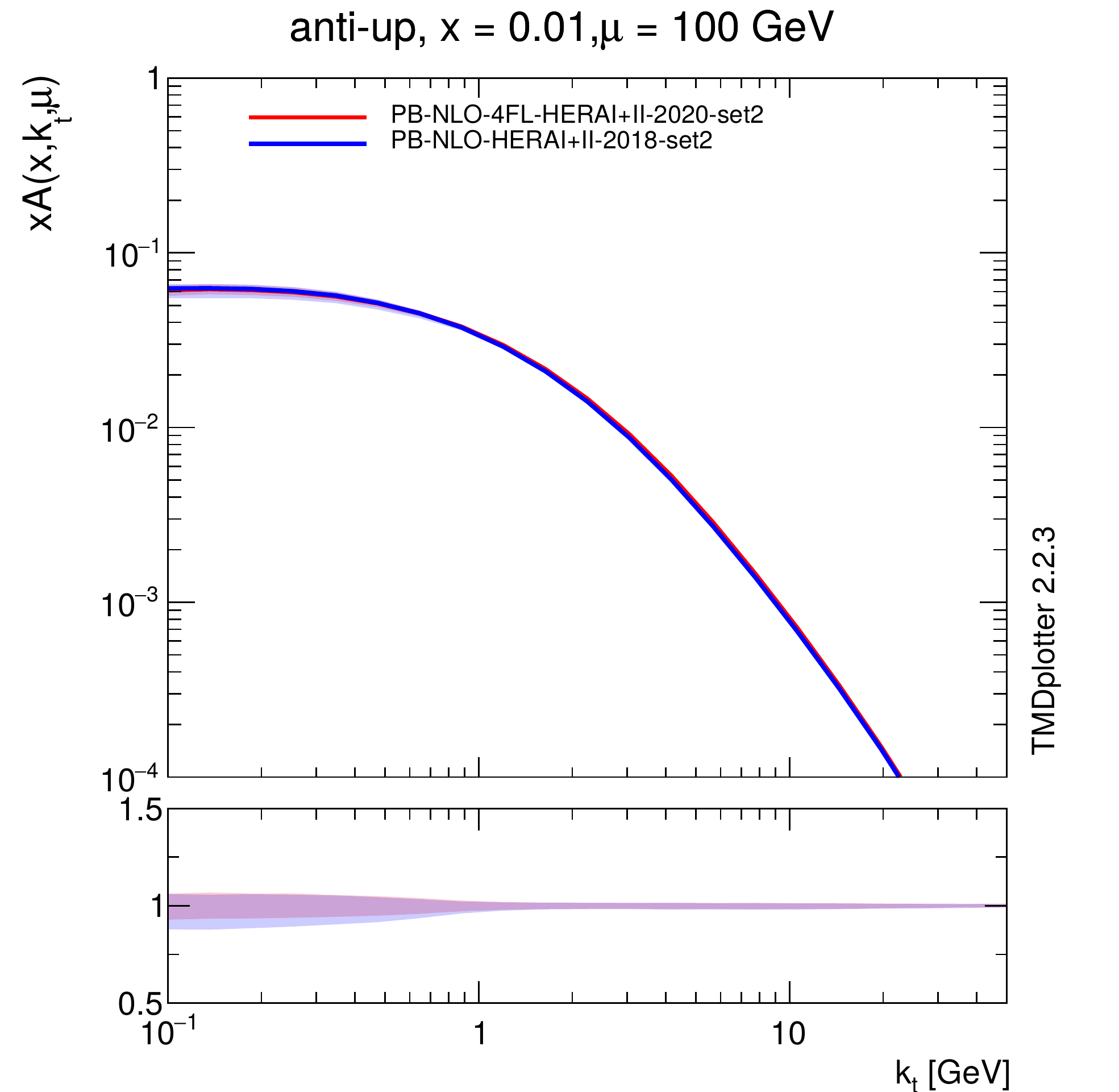}}
\subfigure[charm]{\includegraphics[width=0.32\textwidth]{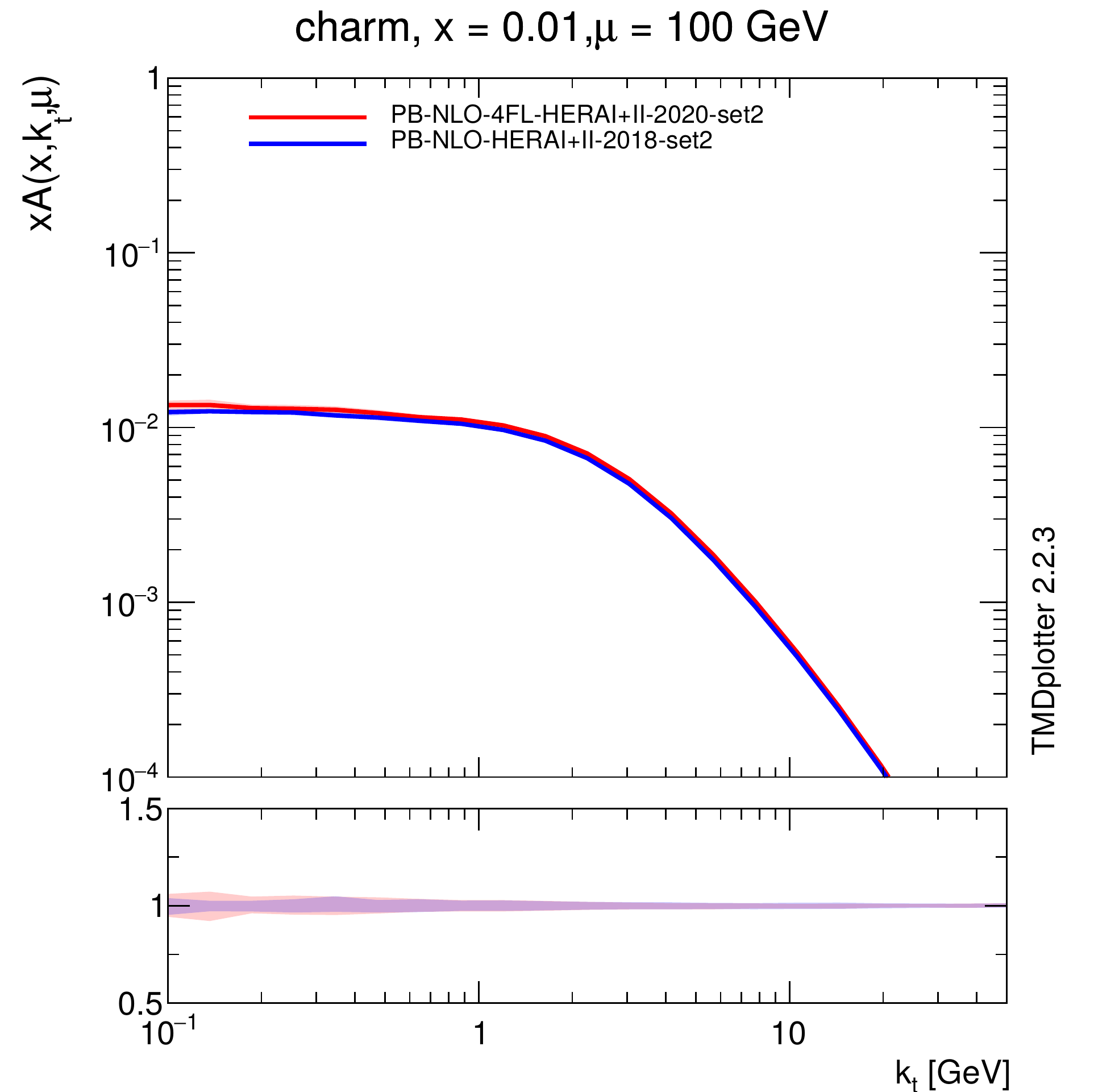}}
\subfigure[bottom and gluon]
{\includegraphics[width=0.32\textwidth]{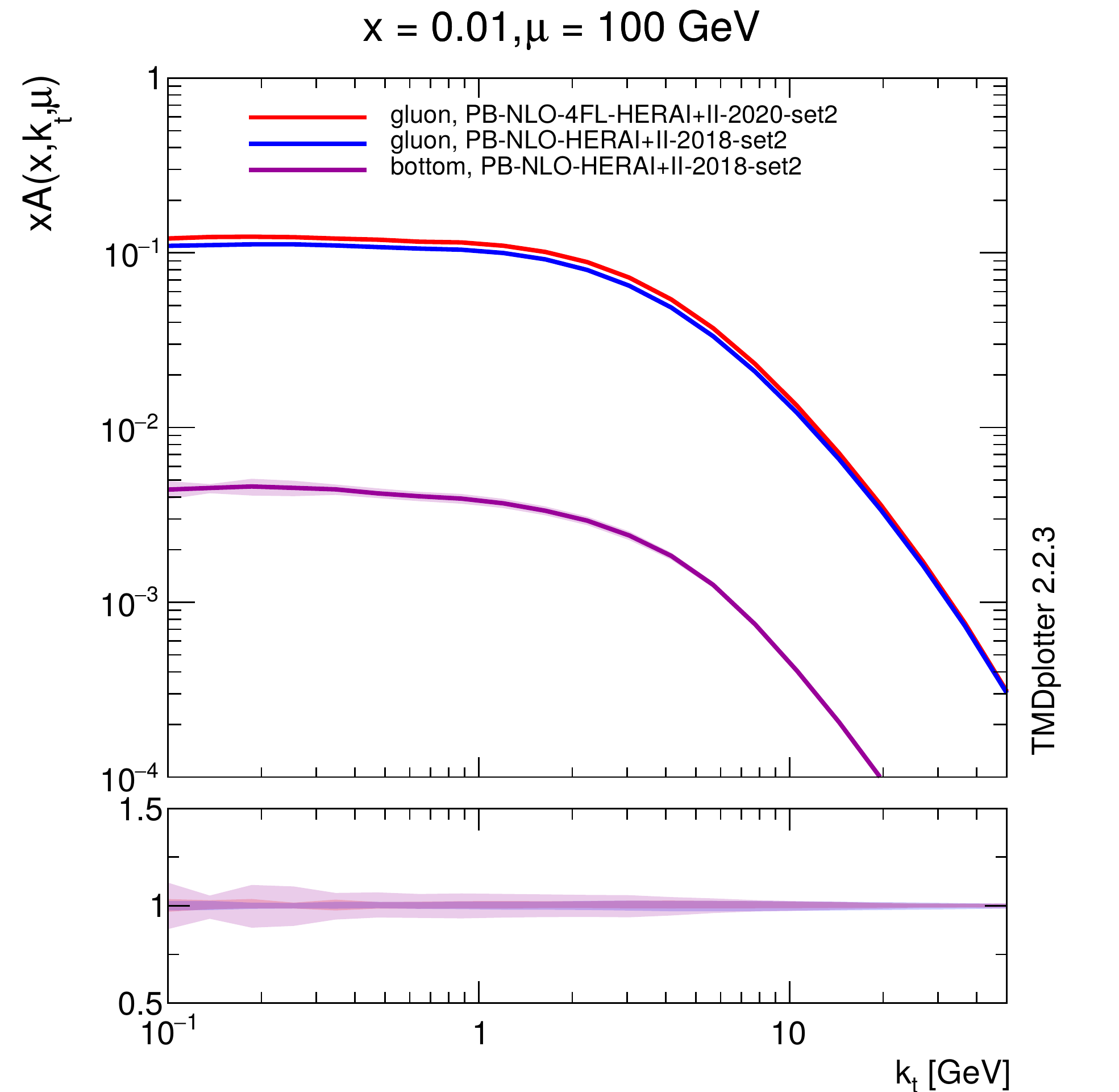}}
\caption{The \Paqu (a), charm (b), gluon and bottom (c) 4 flavour and 5 flavour TMDs as a function of $k_t$ for $x=0.01$ at $\mu=100$ \GeV. }\label{TMDpdfs}
\end{center}
\end{figure}

In Fig. \ref{TMDpdfs} the parton distributions as a function of the transverse momentum \kt\ obtained in the \fourfl\ and \fivefl\ scheme for \Paqu , \Pqc , \Pqb\ and \Pg\ densities at $x=0.01$ and $\mu=100$ GeV (typical values for $Z$ production at the LHC) are shown. The \Pqb -quark exists only in the \fivefl\ scheme, as shown in  Figs.~\ref{pdfs} and Fig.~\ref{TMDpdfs}. 
At small \kt\  essentially the first term in Eq.(\ref{evoleqforA1}) contributes, and the distributions are slightly different because of different magnitudes of the   \fourfl\ and \fivefl\ distributions.  At large \kt\ several branchings may have occurred and the differences between the distributions are washed out. 

In Fig.~\ref{pdfs} and Fig.~\ref{TMDpdfs} the uncertainty band shown corresponds to the uncertainty obtained from the experimental uncertainties, as well as the model uncertainties. 

\section{\boldmath{\PZ + \Pqb}-jet production \label{Sec3}}

The diagrams contributing to \PZ + \Pqb -jet production depend on the scheme used: in the \fourfl -scheme, \Pqb -quarks are produced only in the hard matrix element, while in the \fivefl - scheme, \Pqb -quarks are present already in the parton distribution and thus are treated similar to other lighter quarks. Examples of lowest order diagrams are shown in Fig.~\ref{feynman-d}. 

\begin{tolerant}{1500}
The hard processes are calculated at next-to-leading order (NLO)  with the \MGvATNLO\  package (labelled as MCatNLO), as \PZ + one parton process in the \fivefl -scheme and  $\PZ +\Pqb\Paqb$ in the \fourfl -scheme, using the corresponding \fivefl\ or \fourfl\ collinear parton densities with their different values of $\alphas$. The renormalization and factorization scales are set to $\mu_R = \mu_F = 1/2 \sum_i H_{T,i}$ where $i$ runs over all partons. 
The scale uncertainty is obtained by varying both scales independently by a factor 2 up and down. In the calculation of the hard process with the MC@NLO method~\cite{Frixione:2006gn,Frixione:2003ei,Frixione:2002bd,Frixione:2002ik}, the HERWIG6 subtraction terms are applied, as they are relevant for the use with \PBM -TMDs and the \PBM -TMD parton shower  implemented in   \cascade 3~\cite{Baranov:2021uol}. The transverse momenta of the incoming partons are obtained from the \PBM -TMDs, in the \fivefl\ or \fourfl -scheme, respectively.
\end{tolerant}

\begin{figure}[htb]
\subfigure[] {\includegraphics[width=0.49\textwidth]{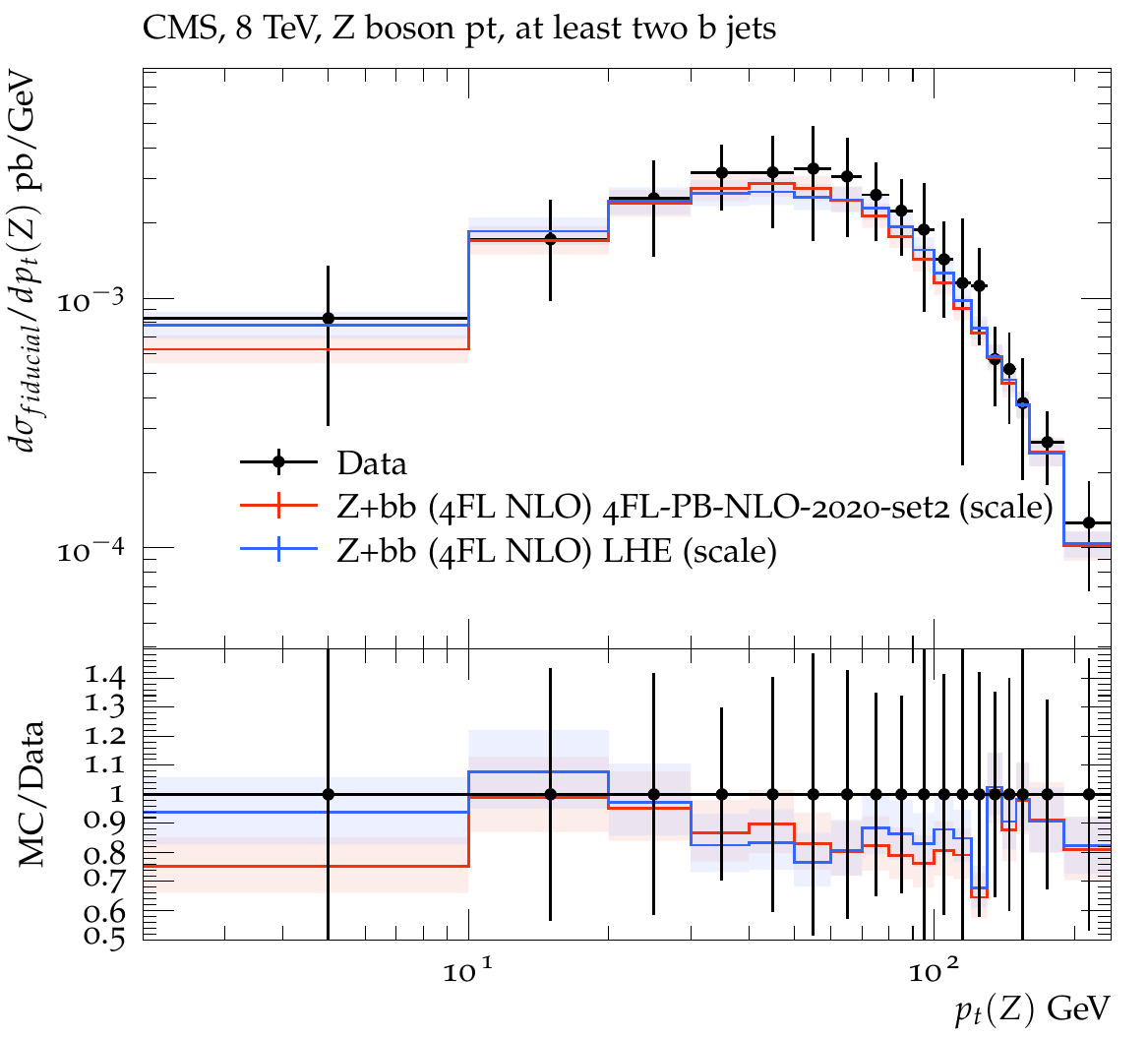}}
\subfigure[]{\includegraphics[width=0.49\textwidth]{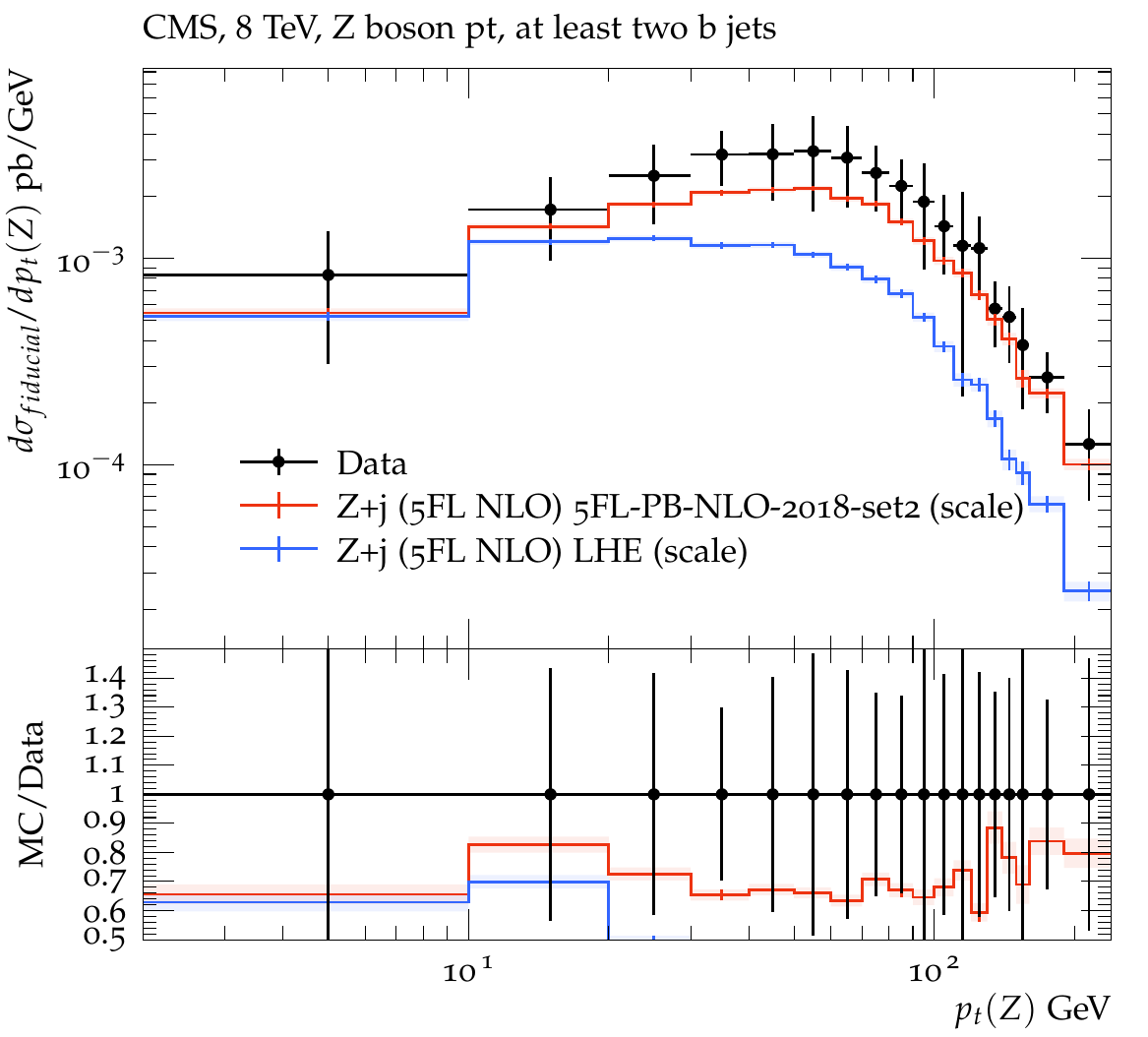}}
\caption{Differential cross section for $\PZ+ \Pqb\Paqb$ tagged jets as a function of the transverse momentum of the \PZ -boson as measured by CMS \protect\cite{Khachatryan:2016iob} at $\sqrt{s}=8$ \TeV. The \fourfl -prediction is shown in (a), the \fivefl -prediction is shown in (b).  In addition to the full prediction the result of using only the LHE files are shown. }
\label{pt-LHE-full}
\end{figure}

Events are selected in the phase space of the CMS measurement~\cite{Khachatryan:2016iob} of a \PZ -boson in the presence of two \Pqb-tagged jets at $\sqrt{s} = 8$~\TeV\ with $ 71<m_{\Pl\Pal} < 111 $ \GeV , and the \Pqb -tagged jets obtained with the  anti-$k_\text{T}$ algorithm~\cite{Cacciari:2008gp}, with a distance parameter of $R = 0.5$ and with $\pt > 30 $ \GeV\ and $|\eta| < 2.4$.
In Fig.\ref{pt-LHE-full} we show the transverse momentum of the \PZ -boson in the presence of two \Pqb-tagged jets at $\sqrt{s} = 8$~\TeV\ obtained from calculations using the \fourfl\ and \fivefl -schemes with MCatNLO and \PBM TMDs and \PBM TMD parton shower. 

In Fig.~\ref{pt-LHE-full} also the predictions from parton level (LHE-level) are shown to illustrate the effect of the \PBM -TMD distribution  on the transverse momentum spectrum of the \PZ -boson. 

\begin{figure}[htb]
\subfigure[] {\includegraphics[width=0.49\textwidth]{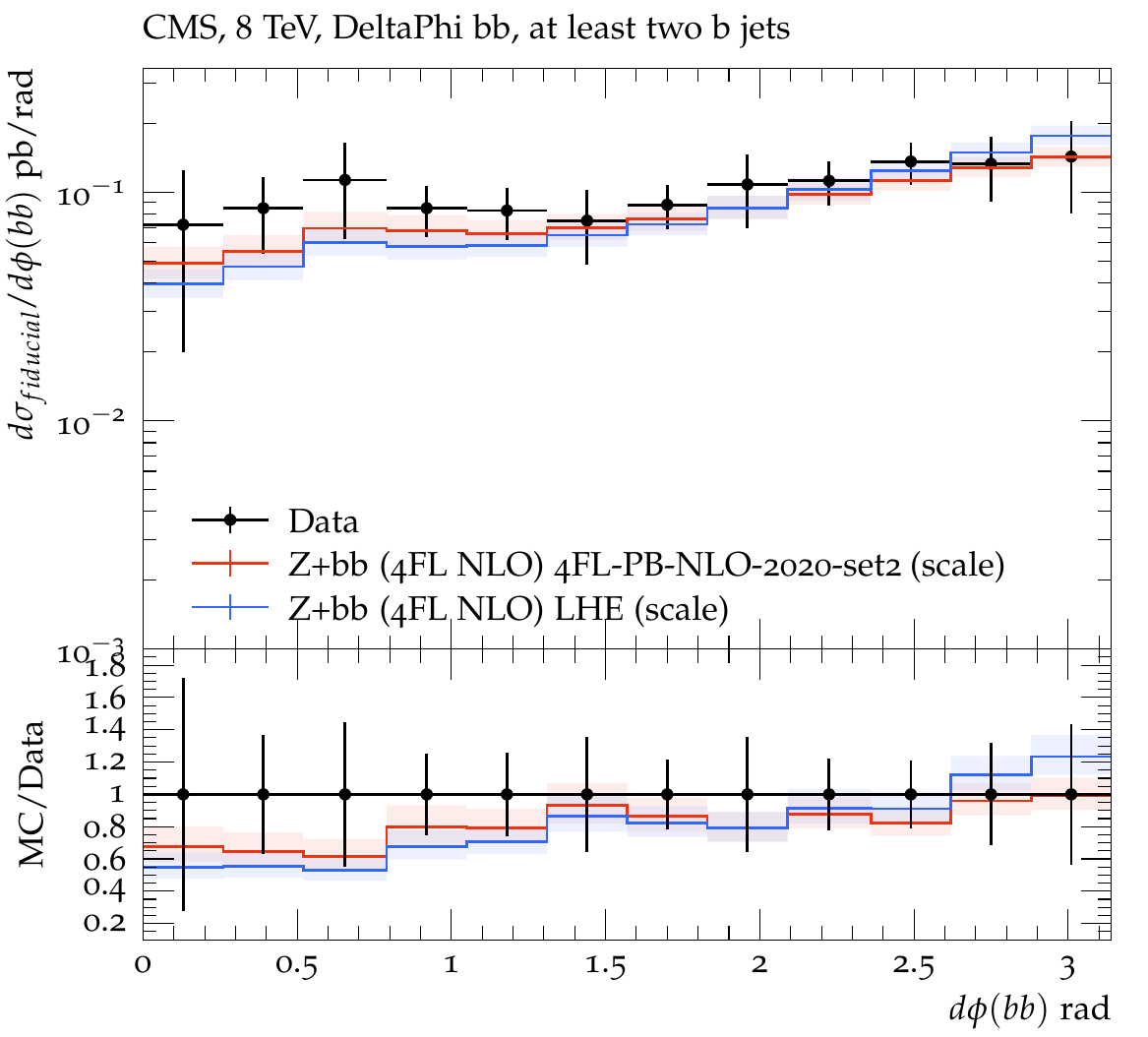}}
\subfigure[]{\includegraphics[width=0.49\textwidth]{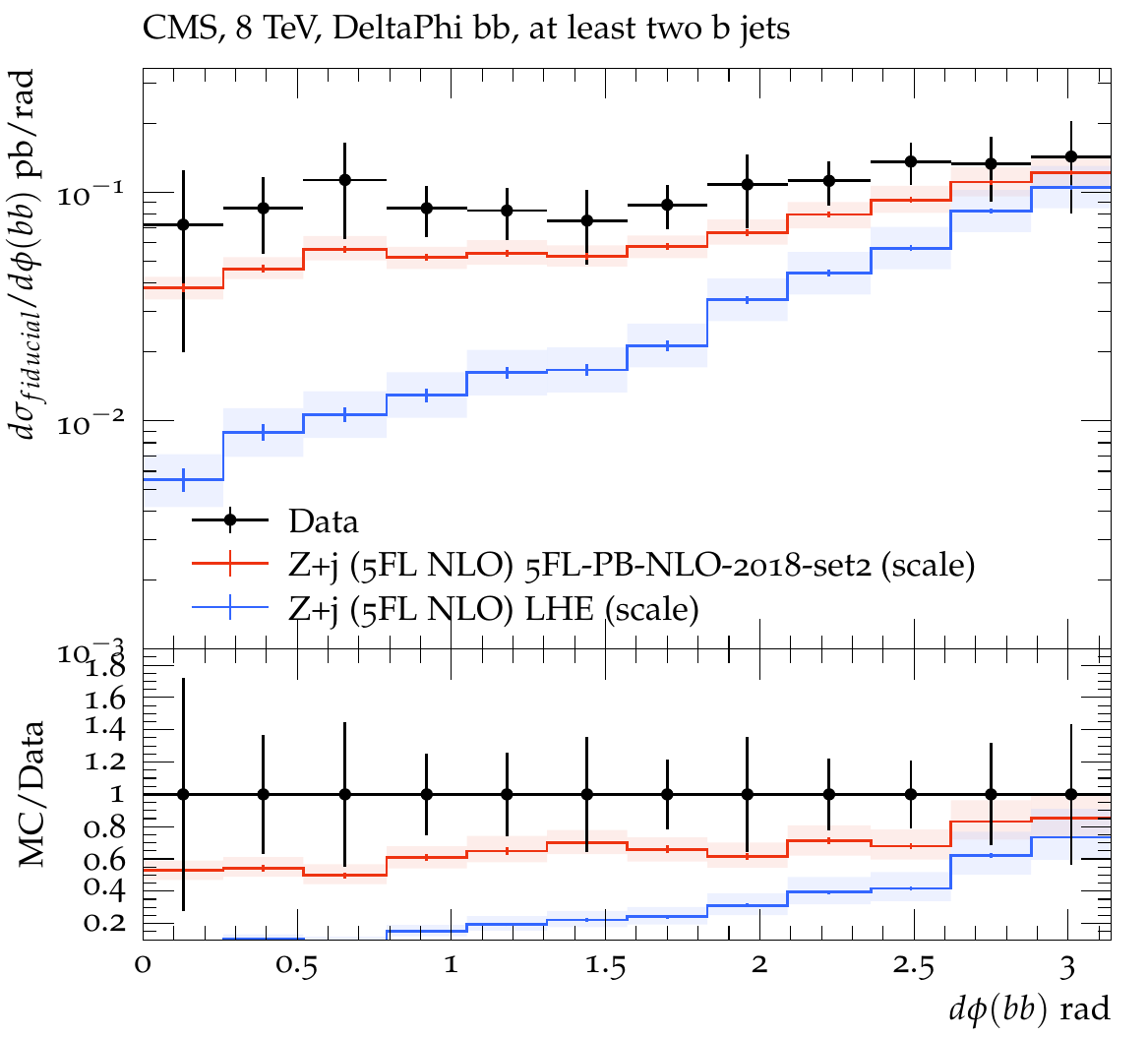}}
\caption{Differential cross section for $\PZ + \Pqb\Paqb$ tagged jets as a function of azimuthal angular separation $\Delta\phi_{\Pqb \Paqb}$ measured by CMS \protect\cite{Khachatryan:2016iob} at $\sqrt{s}=8$ TeV together with predictions at parton level (LHE level) for subtraction terms and after inclusion of PB-TMDs and parton shower in \fourfl\ (a) and \fivefl\ (b) schemes.}
\label{deltaphi-4fl-5fl}
\end{figure}

In Fig.~\ref{deltaphi-4fl-5fl} the cross section as a function of the  azimuthal angular separation between the two \Pqb -tagged jets in $\PZ + \Pqb\Paqb$ events is shown. 
In Fig.~\ref{deltaphi-4fl-5fl}(a) the prediction coming from the \fourfl\ calculation and in Fig.~\ref{deltaphi-4fl-5fl}(b) the prediction using \fivefl\ scheme is compared with the measurement. Also shown are the predictions coming from the pure partonic LHE level, indicating the role of \PBM -TMDs and \PBM -TMD shower. In the \fourfl -calculation, the effect of inclusion of TMDs and TMD shower is small, since both \Pqb\ partons are already produced with NLO accuracy at the matrix element level, while in the \fivefl -scheme, the effect of TMDs and parton shower is large, since a significant contribution comes from \Pqb -quarks inside the parton density and therefore must be simulated in the parton shower. 
The \fourfl\ and \fivefl\ calculations  give very similar results and describe the measurements at a similar level, showing the consistency of both approaches.

\subsection{The role of \PBM -TMD parton shower\label{Sec4}}

\begin{figure}[htb]
\subfigure[] {\includegraphics[width=0.49\textwidth]{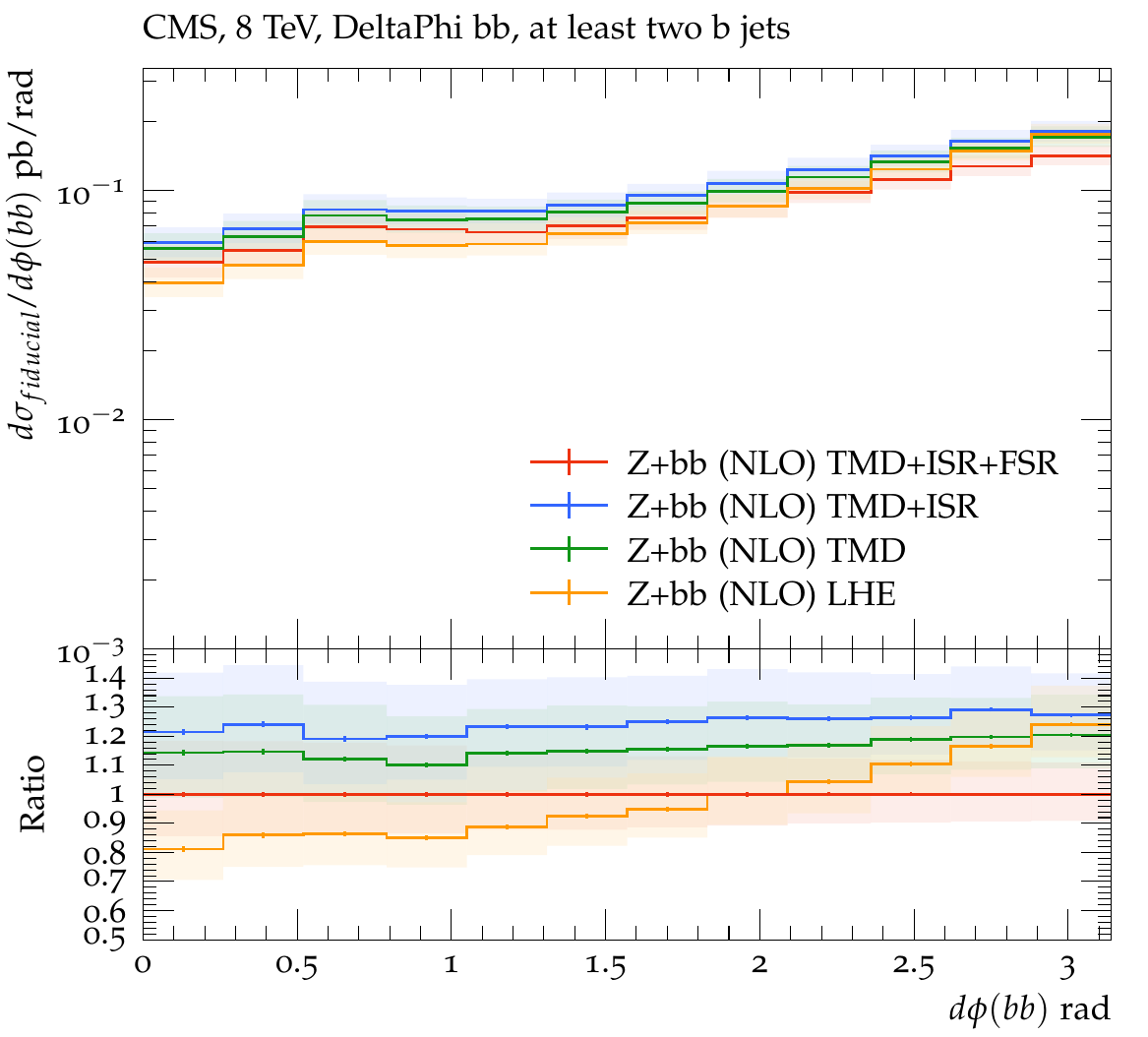}}
\subfigure[]{\includegraphics[width=0.49\textwidth]{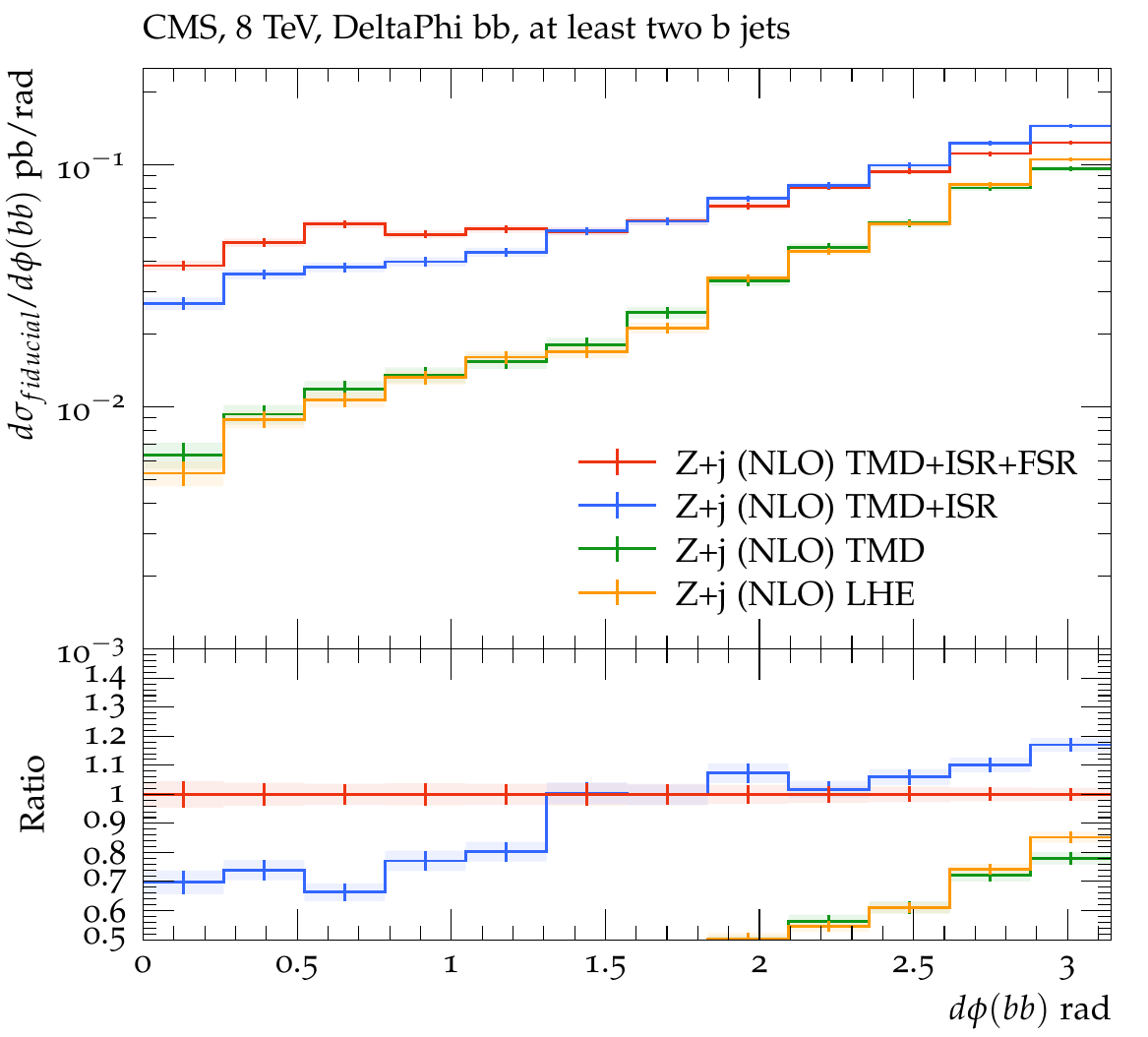}}
\caption{Differential cross section for  \PZ+\Pqb\Paqb tagged jet  production as a function of azimuthal angular separation $\Delta\phi_{\Pqb \Paqb}$ at
 at parton level (LHE level), after inclusion of PB-TMDs, initial state parton shower and final state parton shower.
 In (a) is shown the prediction obtained within the \fourfl -scheme, in (b) the prediction obtained in the \fivefl -scheme.}
  \label{TMD+IPS+FPS}
\end{figure}

In order to quantify the role of \PBM -TMD distributions and the corresponding parton shower on the differential cross section for $\PZ+\Pqb\Paqb$ tagged jets as a function of azimuthal angular separation $\Delta\phi_{\Pqb\Paqb}$, in Fig.~\ref{TMD+IPS+FPS} we show the breakdown of the different contributions, for the \fourfl - and the \fivefl\ calculations separately. 
One can clearly see, that the \fourfl - calculation only weakly depends on \PBM -TMD and parton shower, while a very significant effect is observed for the \fivefl - calculation.  

\begin{figure}[htb]
\centering
\subfigure[] {\includegraphics[width=0.49\textwidth]{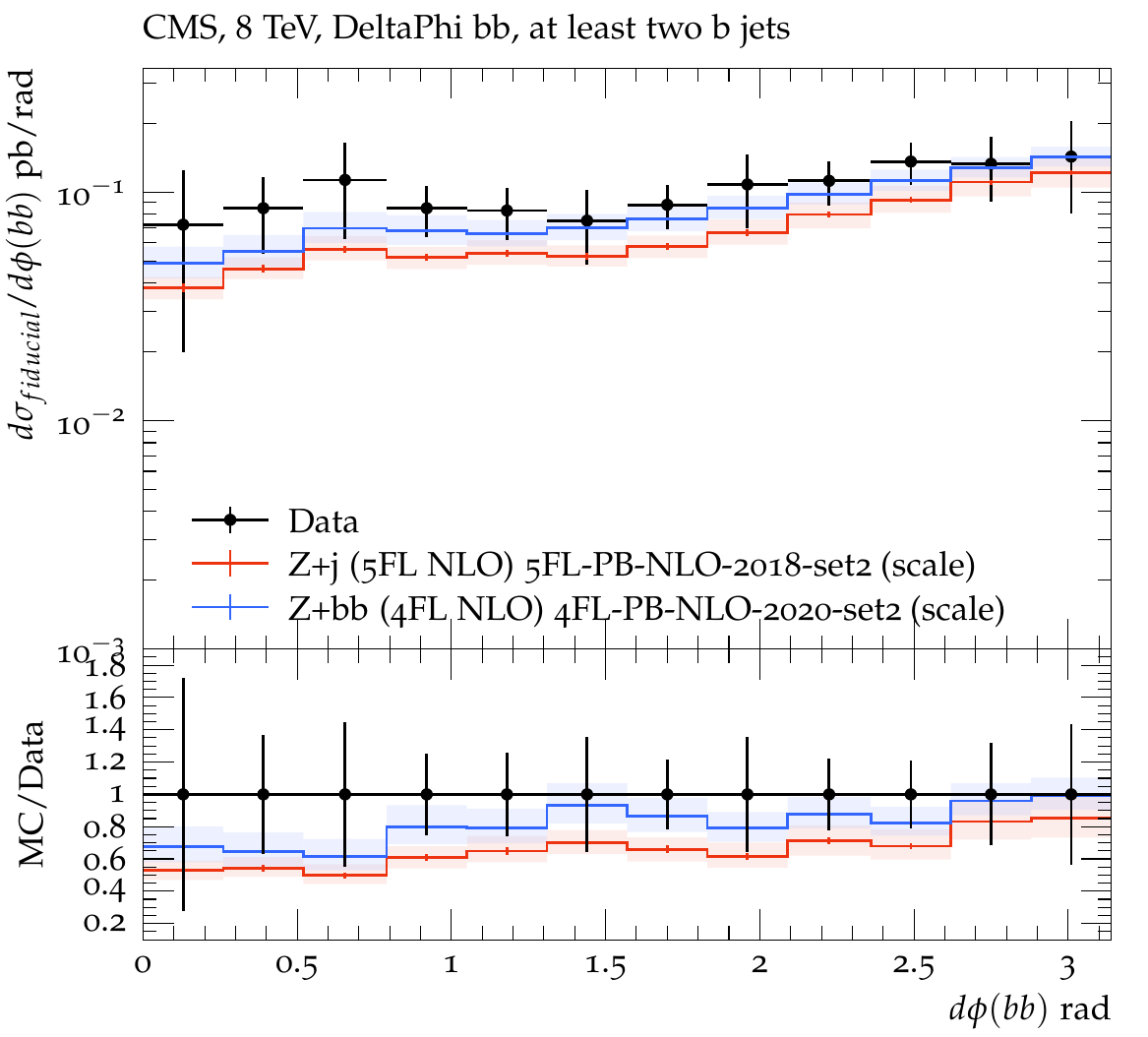}}
\subfigure[]{\includegraphics[width=0.49\textwidth]{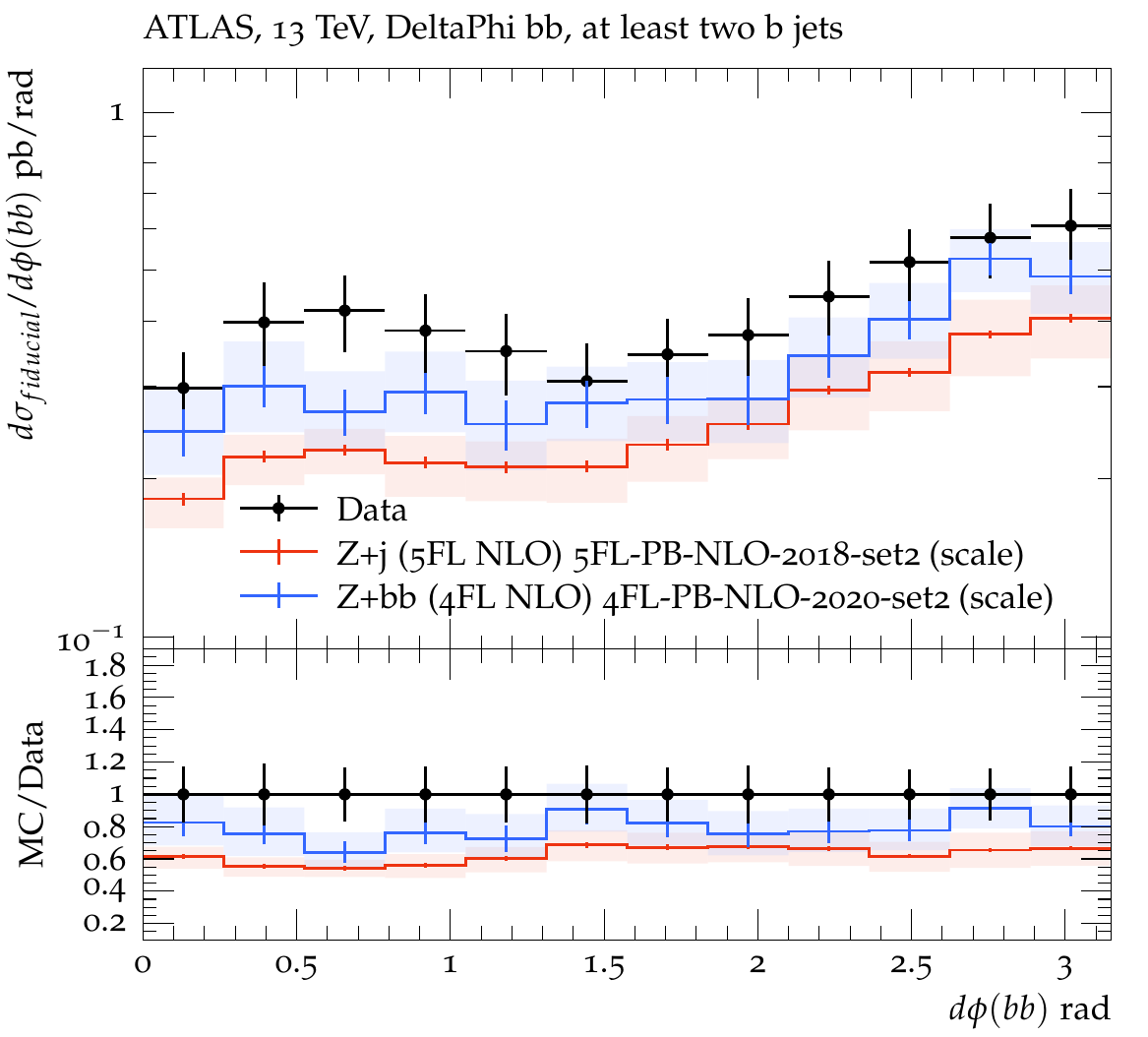}}
\caption{Differential cross section for  $\PZ + \Pqb\Paqb$ tagged jet production as a function of $\Delta\phi_{bb}$ measured by CMS \protect\cite{Khachatryan:2016iob} at $\sqrt{s}=8$ \TeV\ (a) and at 13 \TeV\ (b) as measured by ATLAS~\protect\cite{Aad:2020gfi}.
Shown are the predictions obtained in the \fourfl - and \fivefl - schemes. }\label{deltaphi}
\end{figure}

In Fig.~\ref{deltaphi} we show a comparison of the full prediction obtained in the \fourfl\ and \fivefl -  schemes and compared to a measurement by CMS \protect\cite{Khachatryan:2016iob} at $\sqrt{s}=8$ \TeV\  as well as to a measurement by ATLAS~\cite{Aad:2020gfi}  at $\sqrt{s}=13$ \TeV , showing the very good consistency of both approaches, once the proper parton densities and the corresponding parton shower is included. 
Since the parton distributions are obtained from a fit to HERA data alone, no constraint comes from \Pp\Pp\ or \Pp\Pap\ measurements. However in studies on inclusive jets it is observed that the PB-set2 yields predictions which are in general 10--20 \% below the measurements. Taking a possible shift of 10--20 \% into account, the \PZ+\Pqb\ measurements are very well described.

\section{Conclusion\label{Sec5}}

The calculations of $\PZ + \Pqb\Paqb$ tagged jet production performed in the \fourfl\ - and \fivefl\ - schemes allow for detailed comparison of the heavy flavour structure of collinear and  transverse momentum dependent (TMD) parton distributions as well as for detailed investigations of heavy quarks radiated during the initial state parton shower cascade.

We have determined the first set of collinear and TMD parton distributions in the \fourfl -scheme with NLO DGLAP splitting functions within the \PBM\ approach. The functional form of the initial parton distributions follows the ones of the \fivefl -scheme while the parameters are re-fitted to inclusive deep inelastic scattering measurements from HERA. The \fourfl\ - and \fivefl\ \PBM -TMD distributions were used to calculate $\PZ + \Pqb\Paqb$ tagged jet production at LHC energies. The predictions obtained are in very good agreement with measurements obtained at $\sqrt{s} = 8, 13 $ \TeV\ by the CMS and ATLAS collaborations.

The different configurations of the hard process in the  \fourfl\ - and \fivefl\ schemes allow for a detailed investigation of the performance of heavy flavor collinear and TMD parton distributions and the corresponding initial TMD parton shower. With  consistently obtained  \fourfl\ - and \fivefl\  \PBM -TMD distributions and TMD parton shower, very good agreement of the final cross sections between the two approaches is obtained, which gives confidence in the evolution of the \PBM -TMD parton densities as well as in the \PBM -TMD parton shower.

\section*{Acknowledgments.}
\begin{tolerant}{1500}
We thank Sasha Zenaiev and Francesco Hautmann for many discussions. 
We are grateful for many discussions on \PZ+\Pqb production to S. Baranov, A. Lipatov and M. Malyshev.
STM thanks the Humboldt Foundation for the Georg Forster research fellowship  and gratefully acknowledges support from IPM. 
\end{tolerant}

\bibliographystyle{mybibstyle-new} 
\raggedright 
\bibliography{/Users/jung/Bib/hannes-bib}
\end{document}